\documentclass[journal]{IEEEtran}
\ifCLASSINFOpdf
\else
\fi
\usepackage{mymacros}

\begin{document}
%
\title{Dimensioning spectrum to support ultra-reliable low-latency communication}

\author{
  \IEEEauthorblockN{
    Andr\'e Gomes\IEEEauthorrefmark{1},
    Jacek Kibi\l{}da\IEEEauthorrefmark{1},
    Nicola Marchetti\IEEEauthorrefmark{2}, and 
    Luiz A. DaSilva\IEEEauthorrefmark{1}
  }
  \\\IEEEauthorblockA{
    \IEEEauthorrefmark{1}\textit{Commonwealth Cyber Initiative, Virginia Tech}, USA, E-mail: \{gomesa,jkibilda,ldasilva\}@vt.edu
  }
  \\
  \IEEEauthorblockA{
    \IEEEauthorrefmark{2}\textit{CONNECT, Trinity College Dublin}, Ireland, E-mail: \{nicola.marchetti\}@tcd.ie
  }
}


%


\maketitle


\begin{abstract}
Industry-led initiatives such as the Next G Alliance (NGA) are currently considering how to dimension the spectrum required to support new classes of services envisioned beyond 5G. In particular, support for \ac{URLLC} brings the challenge of how to dimension stochastic wireless networks to meet stringent reliability and latency requirements. Our analysis indicates that the bandwidth needed to meet \ac{URLLC} goals can be in the order of gigahertz, beyond what is available in today's mobile networks. Network densification can ease those bandwidth needs but requires new deployment strategies involving substantially larger numbers of sites. As an alternative, we consider multi-connectivity and multi-operator network sharing as efficient ways to reduce the demand for bandwidth without outright deployment of additional base stations.
\end{abstract}


%
\IEEEpeerreviewmaketitle

\section{Introduction}

Industry-led initiatives such as the Next G Alliance (NGA), an initiative to advance the next generations of wireless networks in North America, are currently considering how to dimension the spectrum to support new classes of services envisioned beyond 5G \cite{6g-roadmap}. As communication services are identified, dimensioning spectrum corresponds to identifying suitable \acp{KPI} and assessing the amount of spectrum needed to meet the service-level requirements. This is key for future licensing, regulatory, and network deployment plans, and can even decide the fate of emerging communication services. In particular, ultra-reliable low-latency communication services pose a challenge of how to dimension the spectrum in stochastic wireless networks to meet stringent reliability and latency requirements. For instance, a typical URLLC requirement specified by the 3GPP is the transmission of 32 bytes within a 1 ms deadline with a 99.999\% success probability, making URLLC sensitive to even rare events that occur with the probability of 0.001\%.

A natural way to compensate for uncertainty in wireless communication, increase reliability, and reduce latency is to provision additional resources, be it in the form of additional antennas (e.g., multi-antenna communication in \cite{urllc-factory}), spectrum (e.g., wider channels in \cite{urllc-factory} and channel hopping in \cite[\S III.B.3]{urllc-improvements}), base stations (e.g., network densification in \cite{URLLC-principles_popovski2018wireless}), or connections (e.g., multi-connectivity in \cite{net-sharing-ieee-tnsm, wolf2018reliable, Interface-diversity_nielsen2018ultra}). Similarly, several works focus on the prioritization of \ac{URLLC} traffic, which is also a form of provisioning additional resources, as more resources are made available to \ac{URLLC} services compared to other types of services (e.g., scheduling prioritization of \ac{URLLC} traffic over eMBB traffic in \cite{scheduling-urllc-emb}). These, and other similar techniques discussed in \cite{urllc-improvements, URLLC-principles_popovski2018wireless}, show that supporting \ac{URLLC} services requires that additional resources be made available. Yet, it remains an open question what level of resources is needed to achieve \ac{URLLC} goals in large-scale mobile networks. 

\begin{figure}[t]
    \centering
    \includegraphics[width=1\linewidth]{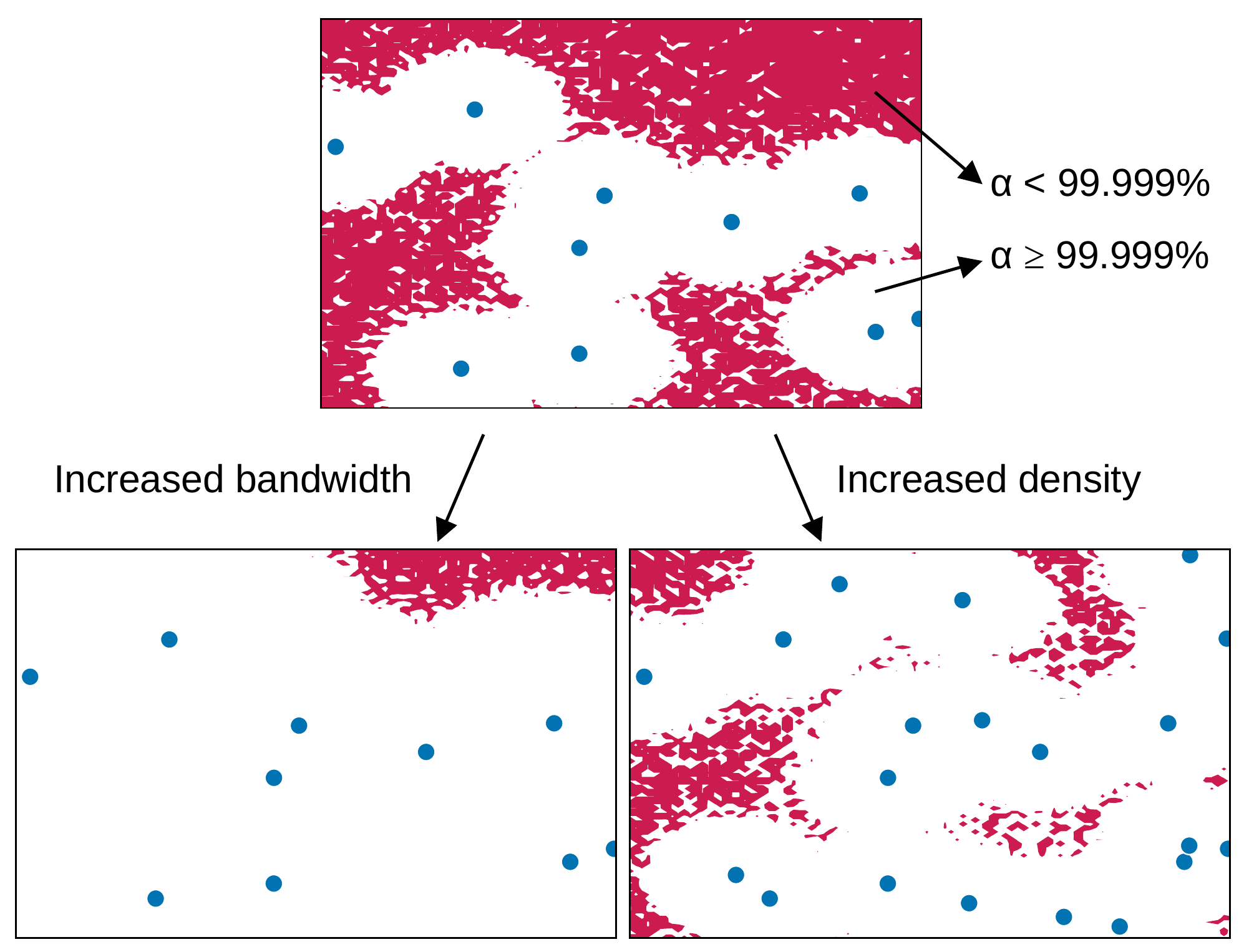}
    \caption{The provision of additional bandwidth and/or network density expands the reach of \ac{URLLC} (white zones, with reliability $\alpha \ge 99.999\%$). Blue markers depict base stations.}
    \label{fig:isobar}
\end{figure}

In this paper, we address this question and study the fundamental limits on the magnitude of resources needed to deploy large-scale mobile networks capable of supporting \ac{URLLC} services. In particular, we seek to determine how to dimension the spectrum. The objective is to guide ongoing and future spectrum standardization efforts as well as future network deployment initiatives regarding the type and magnitude of spectrum required to enable \ac{URLLC} services in mobile networks. To that end, we first map the \ac{URLLC} service-level requirements onto bandwidth requirements and then assess the magnitude and type of spectrum needed to meet the service-level requirements on urban-macro network models developed by the 3GPP. As illustrated in \Fig{isobar}, making more bandwidth available compensates for cell edge effects and expands the reach of \ac{URLLC}. However, our results indicate that supporting \ac{URLLC} services at large-scale networks can demand bandwidth in the order of gigahertz, beyond what is typically available in today's networks. We study how network densification can expand the reach of \ac{URLLC} services without additional bandwidth, as illustrated in the right side of \Fig{isobar}. We then study multi-connectivity and multi-operator network sharing as alternative ways to further ease the demand for bandwidth without outright deployment of additional base stations.

The remainder of this paper is organized as follows. \Sec{preliminaries} describes the mapping between \ac{URLLC} requirements onto bandwidth, followed by a description of the network models considered in this paper. \Sec{results1} presents the numerical results on the magnitude and type of spectrum needed to support \ac{URLLC} services according the the network models developed by the 3GPP. Finally, \Sec{results1} summarizes the main takeaways from this paper. 

\section{Preliminaries} \label{sec:preliminaries}

\subsection{Expressing reliability}\label{sec:reliability}

We adopt the ITU-R/3GPP definition of reliability for \ac{URLLC} \cite{itu-minimum-requirements, 3gpp.38.913}: the success probability $\alpha$ of transmitting $\delta$ bits of data within a user plane latency deadline $\tau$, i.e., $\mathrm{Pr}(T \le \tau) \ge \alpha$. We consider a typical mobile-to-base station link, without intermediate relays. The user plane latency $T$, i.e., the time it takes to transmit $\delta$ bits from/to the base station to/from the mobile in the downlink/uplink communication \cite{3gpp.38.913}, is a sum of physical and data link layers' delays, which generally includes 
\begin{enumerate*}[label=(\alph*)]
    \item a processing delay, incurred by signal processing; 
    \item a medium access delay, incurred by scheduling or contention for the medium; and
    \item a transmission delay, i.e., $T = T_\text{proc} + T_\text{ma} + T_\text{tx}$.
\end{enumerate*} 
The processing delay is fairly constant and hardware/implementation related; the medium access delay can be deterministic (e.g., scheduling-based approaches) or random (e.g., contention-based approaches). In \ac{URLLC}, deterministic medium access is likely to be adopted, given the stringent reliability and latency requirements.

Hence, the transmission delay is the prime source of uncertainty given the probabilistic nature of wireless communication, prone to stochastic channel conditions caused by user mobility, blockages, fading, and interference. 
We adopt a general formulation where the transmission delay relates to the Shannon-Hartley capacity and assume no protocol overhead. This way, we can estimate the \emph{fundamental minimum amount of spectrum} needed to support \ac{URLLC} services. The transmission delay is expressed as:

\begin{equation}
    T_\delta = \frac{\delta}{w\times\log_2(1 + \Gamma)},
    \label{eq:transmit-delay}
\end{equation}

\noindent where $w$ and $\Gamma$ are the bandwidth and \ac{SINR}, respectively. We assume the \ac{SINR} to be static during time $\tau$, which is a reasonable assumption for small values of $\tau$, such as in \ac{URLLC}, where $\tau$ is often 1 \si{\ms}. 

With the focus on the transmission delay, the \ac{URLLC} goals can be expressed as:

\begin{equation}
    \mathrm{Pr}\left(w \ge \frac{\delta}{\tau_\text{tx} \times \log_2(1 + \Gamma)}\right) \ge \alpha,
    \label{eq:reliability}
\end{equation}

\noindent where $\tau_\text{tx}$ is the fraction of the latency budget $\tau$ reserved for the transmission delay. Hence, the minimum bandwidth to satisfy the \ac{URLLC} service-level requirements is a function of the $\alpha$-quantile $Q_{\alpha}$ of the inverse of the spectrum efficiency,

\begin{equation}
    w = \frac{\delta}{\tau_\text{tx}} \times Q_\alpha \left( \frac{1}{\log_2(1 + \Gamma)}\right).
\end{equation}

\subsection{The network model}\label{sec:net-model}

We are interested in the magnitude of spectrum needed to support \ac{URLLC} in a mobile network. To that end, we adopt the 3GPP urban-macro network model in TR 38.901 \cite{3gpp.38.901}. The model defines 
\begin{enumerate*}[label=(\alph*)]
    \item antenna array configurations for base stations \cite[\S 7.3]{3gpp.38.901}; 
    \item path loss \cite[\S 7.4]{3gpp.38.901}; 
    \item fading \cite[\S 7.4]{3gpp.38.901}; and
    \item line-of-sight probabilities \cite[\S 7.4]{3gpp.38.901} for different frequency bands ranging from 0.5 to 100 \si{\GHz}.
\end{enumerate*} 
We consider three frequency bands centered at 700 \si{\MHz}, 4 \si{\GHz}, and 30 \si{\GHz} to study low, mid, and high bands (as recommended by the 3GPP in TR 38.913 \cite{3gpp.38.913}) and base station antenna arrays of size 2x2, 4x4, and 8x8, respectively. Further parameters, in compliance with the aforementioned reports, are as follows:
\begin{enumerate*}[label=(\alph*)]
    \item base stations: 
        \begin{enumerate*}[label=\roman*.]
            \item 25 \si{\meter} high and
            \item 49 \si{dBm} transmit power
        \end{enumerate*}; 
    \item mobile user: 
        \begin{enumerate*}[label=\roman*.]
            \item  1.5 \si{\meter} high,
            \item 0 \si{dBi} omnidirectional antenna, and
            \item 9 \si{dB} noise figure
        \end{enumerate*};
    \item -90 \si{dBm} noise floor.
\end{enumerate*}


\section{What are the minimum spectrum requirements for the mobile network to meet \ac{URLLC} goals?}\label{sec:results1}

In this section, we study the magnitude of spectrum needed to deploy \ac{URLLC}-enabled mobile networks. Let us consider the case of $\delta = 32$ \si{bytes} and $\tau = 1$ \si{\ms} \cite{3gpp.38.913}. We assume processing and medium access delays to be negligible. In practice, processing and medium access delays consume a fraction of the delay budget $\tau$; however, these delays can be considered deterministic in \ac{URLLC}, as discussed in \Sec{reliability}. We focus on the transmission delay, for it involves the primary sources of uncertainty in wireless networks, meaning that the transmission budget takes the entire latency budget, i.e., $\tau_\text{tx} = \tau$ in \Eq{reliability}. This allows us to quantify the \emph{minimum} amount of spectrum needed to support \ac{URLLC} services. We focus on three reliability regimes, which, from now on, we denote as: 
\begin{enumerate*}[label=(\alph*)]
    \item reliable ($\alpha = 90\%$); 
    \item highly-reliable ($\alpha = 99.9\%$); and
    \item ultra-reliable ($\alpha = 99.999\%$).
\end{enumerate*}

We consider two downlink mobile communication scenarios. The first scenario corresponds to mobile networks subject to stochastic path loss and fading, and negligible interference, which we refer to as a noise-limited network, reflecting networks with a low frequency reuse or tight coordination between transmitters. The second scenario, an interference-limited network, includes interference, representing mobile networks subject to full frequency reuse, with little or no interference coordination between transmitters. Practical networks may operate somewhere in between noise- and interference-limited models, for they may adopt some forms of interference mitigation. 
Nevertheless, we consider noise- and interference-limited networks as the \emph{lower}- and \emph{upper}-bound estimates of the \emph{fundamental minimum} amount of spectrum necessary to enable \ac{URLLC} services in mobile networks.

Our results stem from system-level Monte Carlo experiments. In each experiment, we consider the performance of a typical mobile user placed at the origin and base stations deployed according to a \ac{PPP}  
of density $\lambda$ \acp{BS} per \si{\square\km}. The typical mobile associates with the base station of the highest average received power. We assume perfect beam alignment between mobiles and serving base stations and a fixed network load of one active connection per base station. This way, interference at the typical mobile is caused by the beam alignment between non-serving base stations and other randomly placed mobiles in the network. We generate $3\times 10^7$ experiments for each network density to capture even low-probability channel conditions in the order of $0.001\%$ when $\alpha = 99.999\%$.


\subsection{Noise-limited networks}

\begin{figure}[h]
    \centering
    \includegraphics[width=0.97\linewidth]{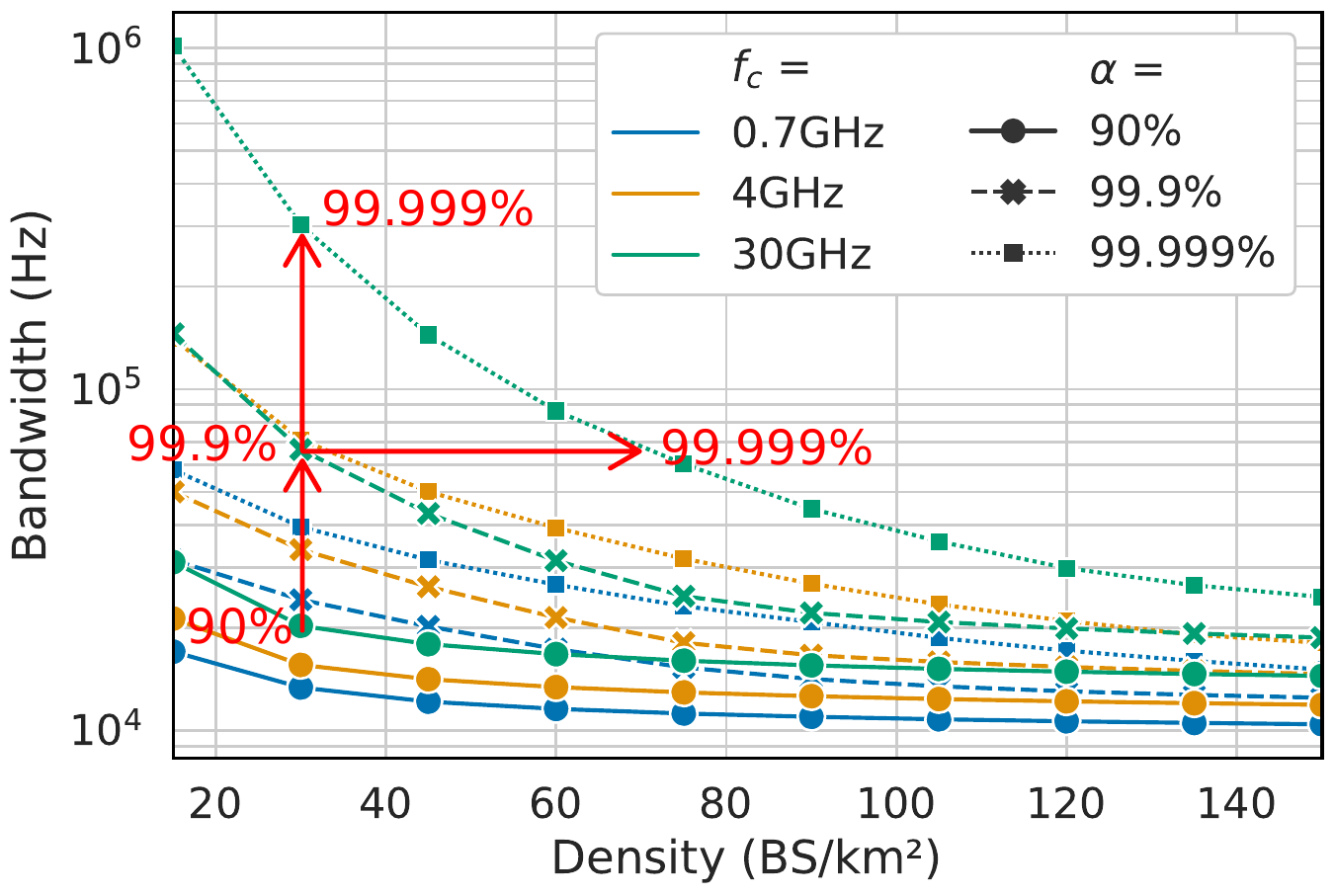}
    \caption{The bandwidth required to meet the \ac{URLLC} requirement of $\delta = 32$ \si{bytes} and $\tau = 1$ \si{\ms} in noise-limited networks operating in different frequency bands.}
    \label{fig:band-density-SNR-3GPP}
\end{figure}

\Fig{band-density-SNR-3GPP} shows the demand for bandwidth to reach certain levels of reliability in different network densities for the three frequency bands of interest, 700 \si{\MHz}, 4 \si{\GHz}, and 30 \si{\GHz}. 
Let us consider a mobile network of density 30 \si{BS\per\square\km}, a density analogous to dense urban networks in \cite{3gpp.38.913}, at the 30 \si{\GHz} band as a reference. This network can be tailored to support reliable (i.e., $\alpha = 90\%$) to ultra-reliable communication (i.e., $\alpha = 99.999\%$) by the provision of an order of magnitude additional bandwidth (see vertical red arrows in \Fig{band-density-SNR-3GPP}). Alternatively, increased density can ease the demand for bandwidth, such as illustrated by the horizontal red arrow in \Fig{band-density-SNR-3GPP}, where ultra-reliable communication demands as much as bandwidth as highly-reliable communication (i.e., $\alpha = 99.9\%$) at the cost of a roughly $2.3\times$ denser deployment.

The relationship between bandwidth and network density changes for different reliability levels $\alpha$. An increase in network density can be traded for up to an order of magnitude of bandwidth in ultra-reliable communication (dotted lines), whereas it has marginal impact on less reliable scenarios, such as $\alpha = 90\%$ (continuous lines). 
This is a consequence of how network density impacts coverage. As illustrated in \Fig{isobar}, the farther the mobile is from the base station, the harder it is to meet the reliability requirements because of weak coverage, increasing the demand for additional bandwidth. Increased network density improves coverage by reducing the distance between transmitter and receiver, as well as the probability of non-line-of-sight communication. As ultra-reliable communication is more sensitive to long distance communication than less reliable regimes, ultra-reliable communication benefits the most from increased network density.


\subsection{Interference-limited networks}\label{sec:results1-interference}

\begin{figure}[h]
    \centering
    \includegraphics[width=0.97\linewidth]{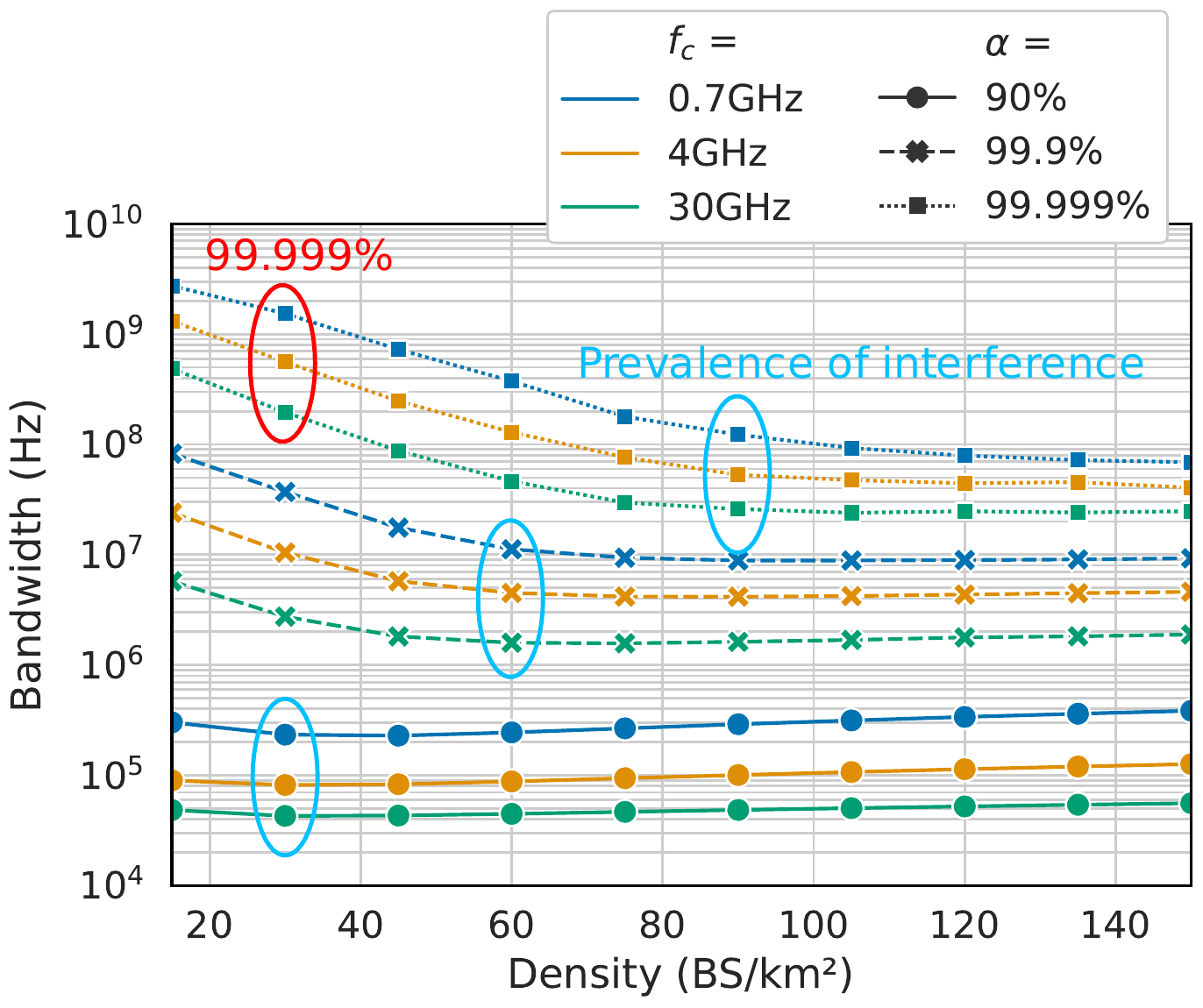}
    \caption{The bandwidth required to meet the \ac{URLLC} requirement of $\delta = 32$ \si{bytes} and $\tau = 1$ \si{\ms} in interference-limited networks operating in different frequency bands. The red circle indicates the bandwidth required for ultra-reliable communication when the network density is 30 \si{BS\per\square\km}. Blue circles indicate the network density where interference prevails for different reliability levels.}
    \label{fig:band-density-SINR-3GPP}
\end{figure}

Interference-limited networks demand additional spectrum to compensate for interference. \Fig{band-density-SINR-3GPP} is \Fig{band-density-SNR-3GPP}'s counterpart and shows that achieving highly- (i.e, $\alpha = 99.9\%$) and ultra-reliable communication (i.e., $\alpha = 99.999\%$) can demand orders of magnitude more bandwidth than in noise-limited networks, whereas the requirements are somewhat similar for reliable communication (i.e., $\alpha = 90\%$) in both networks. Our system model assumes highly directional antennas with perfect beam alignment, which naturally decreases the overall interference, being it marginal when $\alpha \le 90\%$. However, as the reliability requirement increases, communication is more sensitive to even low-probability interference conditions, such as in edge-case scenarios when nearby non-serving base stations (i.e., interferers) steer their beams towards the mobile, demanding wider bandwidth to compensate for strong interference.

The amount of bandwidth to support ultra-reliable communication is beyond what is typically available in today's networks. In dense urban networks, where network density is approximately 30 \si{BS\per\square\km} \cite{3gpp.38.913}, the required bandwidth to meet ultra-reliable communication is around a gigahertz, which is impossible in the 700 \si{\MHz} band and impractical in the 4 \si{\GHz} band. The substantial amount of bandwidth favors the use of higher frequency bands, where wider bandwidth is typically available. Furthermore, higher frequency bands are often coupled with denser antenna arrays (as in our network model), enabling higher directionality transmission that reduces interference and increases resilience to path loss. This translates into less spectrum required to support \ac{URLLC} services than in the case of lower frequency bands. 

Increased network density can ease the bandwidth requirements to some extent. Network density increases coverage until interference prevails, that is, dominant interfering links shift from non-line-of-sight to line-of-sight, and network densification starts to have a negligible (or even harmful) effect on coverage \cite{los-nlos}. Interestingly, interference prevails at different network densities for different reliability levels. For instance, \Fig{band-density-SINR-3GPP} suggests that increased network density leads to marginal returns in reliable communication, where interference prevails at approximately 30 \si{BS\per\square\km}. On the other hand, densification substantially eases the demand for bandwidth in highly- and ultra-reliable communication up to the prevalence of interference at approximately 60 \si{BS\per\square\km} and 90 \si{BS\per\square\km}, respectively. 

\begin{figure}[h]
    \centering
    \includegraphics[width=0.97\linewidth]{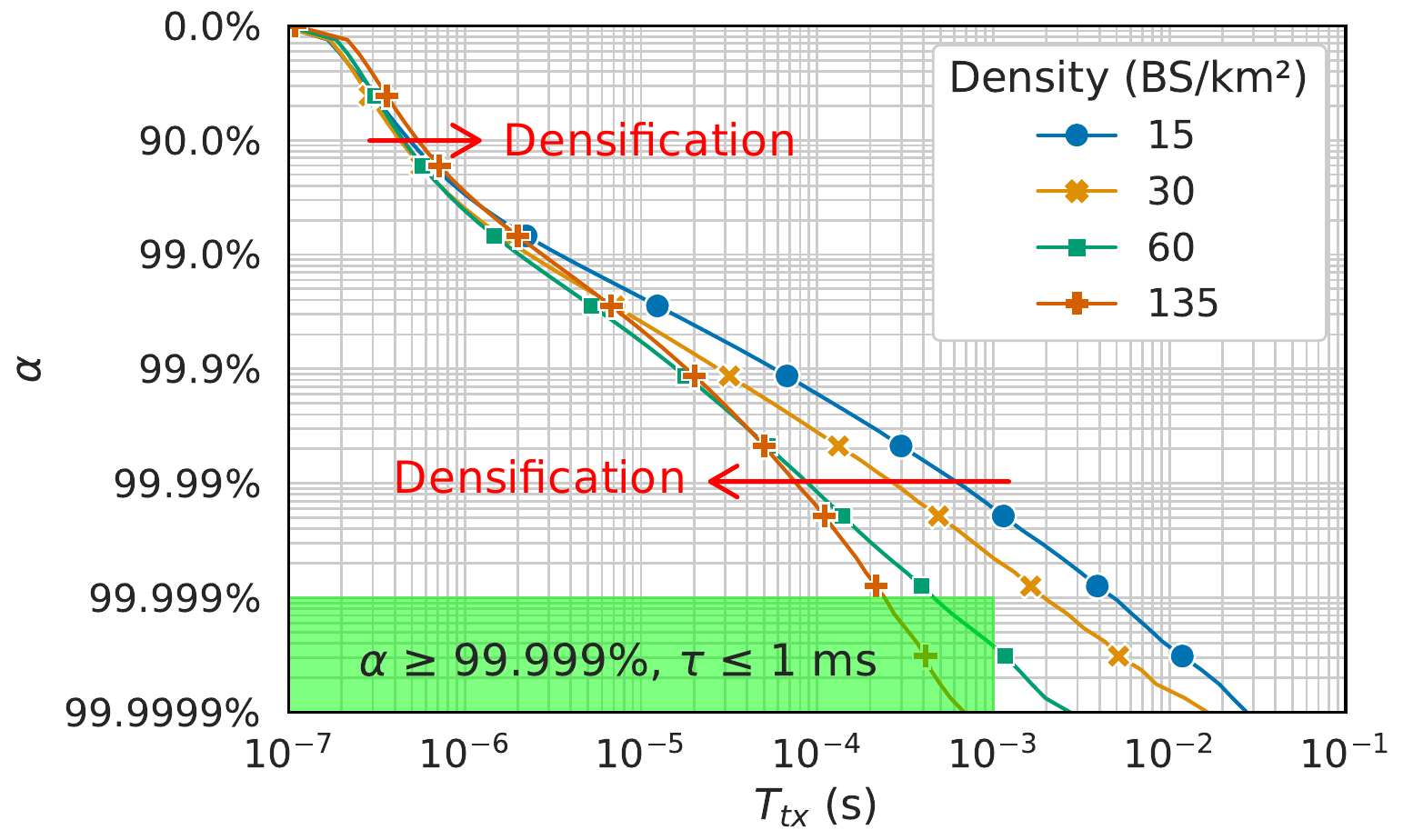}
    \caption{The impact of network density on the transmission delay when bandwidth is fixed at 100 \si{\MHz} in the 30 \si{\GHz} band. A general \ac{URLLC} requirement of $\alpha = 99.999\%$ and $\tau = 1$ \si{\ms} (green zone) demands densities $\ge 60$ \si{\ac{BS}\per\km\square}. Similar patterns were found for the 700 \si{\MHz} and 4 \si{\GHz} bands.}
    \label{fig:cdf-delay-30ghz}
\end{figure}

Another way to look at the impact of network density is to consider the delay at a fixed bandwidth. \Fig{cdf-delay-30ghz} illustrates the mapping from $\alpha$ and $T_\text{tx}$ to network density when bandwidth is fixed at 100 \si{\MHz} in the 30 \si{\GHz} band, the same amount of bandwidth recently auctioned in the US in auction 101 of the Federal Communications Commission (FCC) for one block of spectrum in upper 30 \si{\GHz} bands.
Lower reliability regimes (e.g., $\alpha \le 99\%$) are only marginally impacted by network density; in fact, densification tends to increase the transmission delay in that region, as shown in the upper left side \Fig{cdf-delay-30ghz}. On the other hand, as we increase the reliability requirements (e.g., $\alpha \ge 99.999\%$), there is a substantial demand for increased network density. For instance, a general \ac{URLLC} requirement of $\alpha = 99.999\%$ and $\tau = 1$ \si{\ms} demands a network density of 60 \si{\ac{BS}\per\km\square} or denser in \Fig{cdf-delay-30ghz}. 

\begin{figure}[h]
    \centering
    \includegraphics[width=0.97\linewidth]{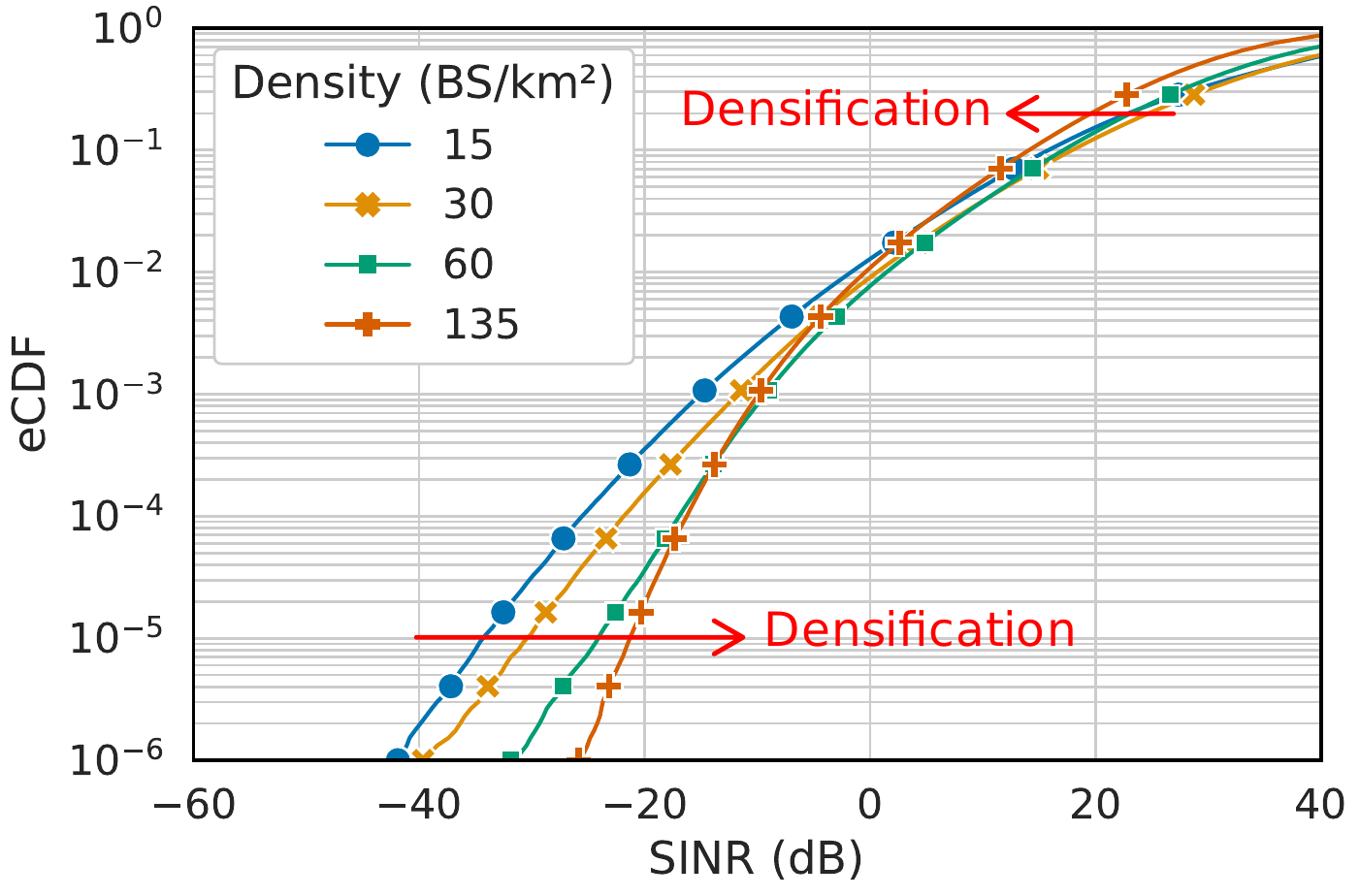}
    \caption{The empirical cumulative distribution function of the \ac{SINR} for different densities in the 30 \si{\GHz} band. Similar patterns were found for the 700 \si{\MHz} and 4 \si{\GHz} bands.}
    \label{fig:cdf-sinr}
\end{figure}

The distinct trends in network density for different reliability regimes are due to how network density impacts overall and edge coverage. Unlike in noise-limited networks, where network density directly maps onto signal strength, the relationship between network density and coverage is subtle in interference-limited networks, for increased network density increases both signal strength and interference. As shown in \Fig{cdf-sinr}, increased network density incurs a small penalty for the overall coverage (e.g., coverage probability $\ge 10^{-2}$, or 1\%); on the other hand, increased network density significantly increases edge coverage (e.g., coverage probability $< 10^{-2}$). (Results consistent with this finding are reported in \cite{noise-interference-limited-regimes}).  
The sensitivity to highly improbable events, such as edge conditions, increases with the reliability degree $\alpha$ (e.g., events with probability as low as 0.001\% when $\alpha = 99.999\%$). As such, ultra-reliable communication is more sensitive to edge coverage than less reliable regimes, benefiting the most from increased network density. 

\subsection{Easing the bandwidth requirements}

As discussed in the previous section, deploying mobile networks capable of supporting ultra-reliable communication ($\alpha \ge 99.999\%$) can demand massive amounts of resources (e.g., $\ge 1$ \si{\GHz} bandwidth or $\approx 100$ \si{\ac{BS}\per\square\km}). In this section, we consider multi-connectivity and multi-operator connectivity as strategies to reduce this  demand. 

Multi-connectivity is an enabler of \ac{URLLC} that leverages diversity and redundancy in the form of multiple connections to mitigate the volatility of a wireless link \cite{urllc-improvements, wolf2018reliable, Interface-diversity_nielsen2018ultra}. In multi-connectivity, a mobile multi-connects to distinct base stations. The received signals can be combined in several ways using diversity combining techniques. In our analysis, we assume signals are weighted and summed according to the respective signal strengths, as in maximum ratio combining.

Multi-operator connectivity combines multi-connectivity and multi-operator network sharing. In multi-operator connectivity, the mobile can multi-connect to base stations of different mobile operators in a multi-operator network. The multi-operator network sharing arrangement is akin to mobile virtual network operators like Google Fi, in the US, which operates atop several mobile operators. However, unlike Google Fi-like multi-operator networks, where users are often subject to the same radio resource management policy as subscribers of the underlying operators, we assume that bandwidth can be reserved and isolated for \ac{URLLC} users at each operator, for instance through network slicing. Furthermore, we assume users can simultaneously connect to base stations of more than one operators, as opposed to a single connection to a single base station at a time.

The resources in multi-operator networks differ from traditional single-operator networks in two ways. First, the base stations exhibit spatial correlation because of clustering between the involved operators \cite{multi-operator-bs-placement}. Intuitively, clustering is caused by similar network planning strategies by the operators, such as covering areas of mutual interest, such as dense central business districts, hence deploying base stations adjacent to each other, or sharing towers, hence co-locating infrastructure. Second, multi-operator connectivity accounts for greater spectrum diversity because each mobile operator operates in its own licensed spectrum. 

\begin{figure}[h]
    \centering
    \includegraphics[width=0.9\linewidth]{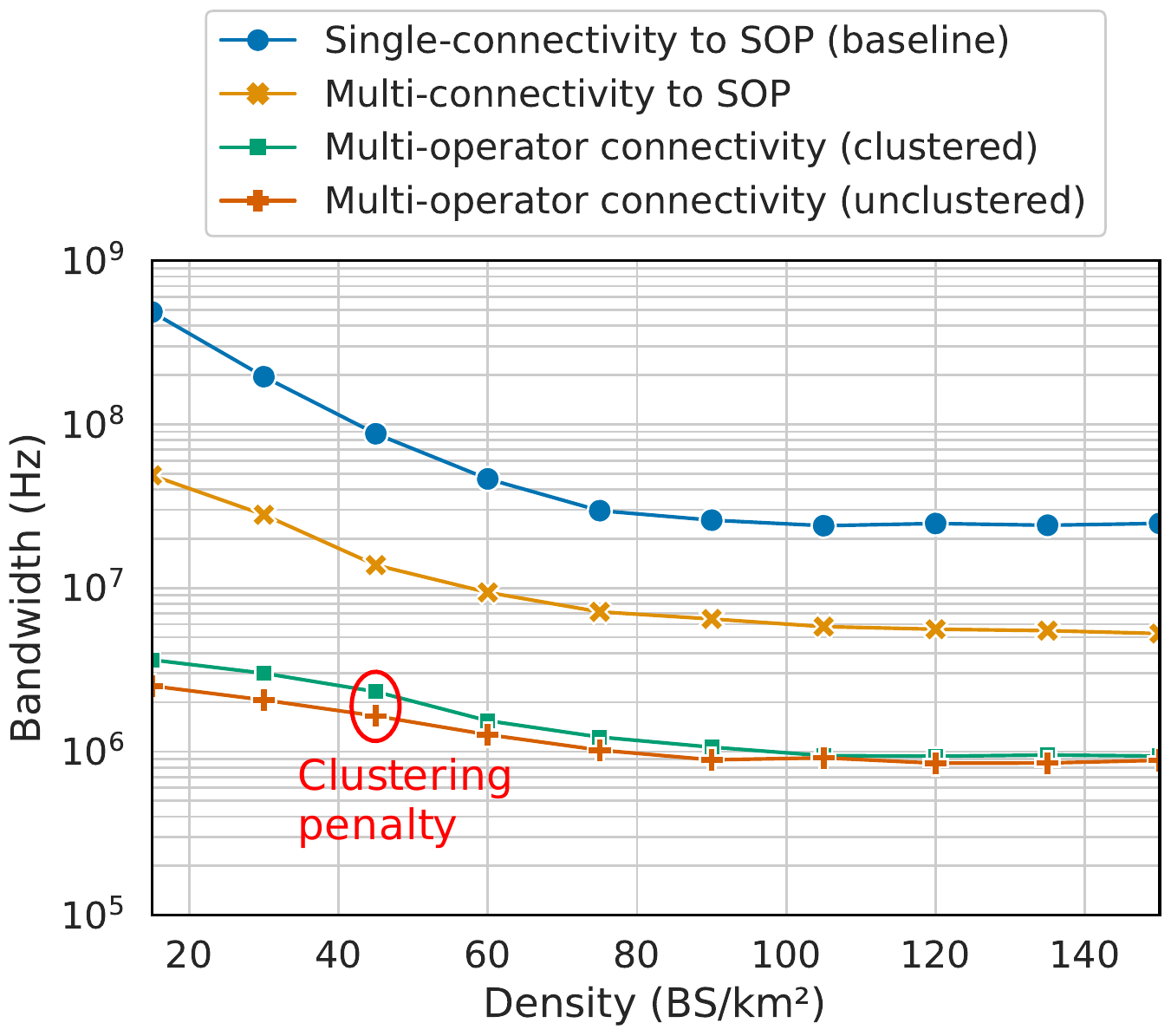}
    \caption{The bandwidth required to meet the \ac{URLLC} requirement of $\delta = 32$ \si{bytes}, $\tau = 1$ \si{\ms}, and $\alpha = 99.999\%$ at the 30 \si{\GHz} band in interference-limited networks. SOP stands for single-operator networks. The multi-operator network in multi-operator connectivity consists of two operators.}
    \label{fig:mop}
\end{figure}

As in \Sec{results1-interference}, we consider interference-limited networks. \Fig{mop} shows the bandwidth required to meet the \ac{URLLC} requirement of $\delta = 32$ \si{bytes}, $\tau = 1$ \si{\ms}, and $\alpha = 99.999\%$. We focus our analysis on the 30 \si{\GHz} frequency band; nevertheless, we can report that similar patterns were found in the 700 \si{\MHz} and 4 \si{\GHz} bands. Here, single-connectivity consists of a single connection to a single operator (as in previous sections; hence our baseline); multi-connectivity, two connections, each to a base station of the same operator; and multi-operator connectivity, two connections, one connection to a base station of operator A and one connection to a base station of operator B. We model clustering in multi-operator connectivity as a Gauss-Poisson point process where clusters are deployed according to a \ac{PPP} of density $\lambda/2$ \cite{gpp}. Each cluster consists of two points: the first is at the center of the cluster and corresponds to a base station of operator A, and the second is randomly placed within a radius of 50 \si{\meter} and corresponds to a base station of operator B. Each mobile operator operates in its own licensed band with full frequency reuse.

Multi-connectivity reduces the required bandwidth by approximately one order of magnitude. The reduction in the amount of bandwidth stems from the spatial diversity introduced by simultaneous connections to two base stations, which results in connections with distinct line-of-sight, path loss, and fading patterns. On the downside, both connections are subject to the same interference component because of full frequency reuse by the operator. 

Multi-operator connectivity, in turn, benefits from frequency diversity, as each operator operates in its own licensed spectrum. As a result, multi-operator connectivity further reduces the required bandwidth, approximately two orders of magnitude compared to the baseline, from hundreds of megahertz to a few megahertz when the network density is 30 \si{BS\per\square\km}. Clustering incurs a small penalty in multi-operator connectivity compared to what it would be if the operators were unclustered (i.e., independent deployments), marginally increasing the demand for bandwidth. Furthermore, multi-operator connectivity is somewhat insensitive to the network density compared to single-connectivity and multi-connectivity and does not require massive network densification to reduce the demand for bandwidth, hence potentially reducing the cost of deploying networks capable of supporting \ac{URLLC} services.

\section{Conclusion}\label{sec:conclusion}

Dimensioning spectrum is key to ongoing and future spectrum standardization efforts as well as future network deployment campaigns to enable \ac{URLLC} services. In this paper, we studied how to dimension the spectrum for \ac{URLLC} services by first mapping the \ac{URLLC} requirements onto bandwidth and then assessing the amount of bandwidth it takes to support \ac{URLLC} services in large-scale 3GPP mobile networks. Our study indicates that the demand for bandwidth heavily depends on the reliability requirement of interest. In ultra-reliable communication, where reliability is $\ge 99.999\%$, the amount of bandwidth can be in the order gigahertz, as opposed to less than a megahertz when reliability is $90\%$. Higher frequency bands, often coupled with wider bandwidth and denser antenna arrays than lower bands, are required in order to meet ultra-reliability goals. The demand for bandwidth can be traded for network density, but it requires massive densification (e.g., $\approx 100$ \si{BS\per\square\km}) to significantly lower the required bandwidth, which is costly or even impractical, depending on the spatial availability for deployment of base stations. Alternatively, multi-operator connectivity, a form of multi-connectivity to multiple mobile operators, was shown to reduce the demand for bandwidth by a few orders of magnitude without the outright deployment of a massive number of base stations.

As future work, we plan to investigate other sources of randomness in the network, such as network load, and ways to circumvent their impact on the demand for network resources in the context of \ac{URLLC}, such as using traffic prediction techniques. As a byproduct of this paper, we also plan to study the problem of efficiently estimating stringent reliability \acp{KPI}. As in this paper, estimating the bandwidth at the 99.999\% reliability level is time-consuming because it takes $> 10^5$ data samples (or simulation cycles) to capture rare network conditions that may occur only 0.001\% of the time, which can be impractical depending on the time it takes to collect each sample and the time window available for data collection and assessment.

\section*{Acknowledgments}
The research leading to this paper received support from the Commonwealth Cyber Initiative (CCI) in Virginia, the US, an investment in the advancement of cyber R\&D, innovation, and workforce development: \url{www.cyberinitiative.org}. It was also supported by the Science Foundation Ireland under grants 17/NSFC/5224 and 13/RC/2077\_P2 (CONNECT). 



%

\bibliography{IEEEabrv, main.bib}
\bibliographystyle{IEEEtran}

\begin{IEEEbiographynophoto}
{Andr\'e Gomes} is a Ph.D. student in computer engineering with the Commonwealth Cyber Initiative (CCI) at Virginia Tech. 
He was previously with the Irish National Telecommunications Research Centre (CONNECT), Trinity College Dublin, Ireland. 
He received a B.Sc. degree in telecommunications engineering from the Universidade Federal de São João del-Rei (UFSJ), Brazil, was a visiting student at the University of Adelaide, Australia, and holds an M.Sc. degree in computer science from the Universidade Federal de Minas Gerais (UFMG), Brazil. His research interests include network reliability, self-organising networks, and software-defined networks.
\end{IEEEbiographynophoto}

\begin{IEEEbiographynophoto}{Jacek Kibi\l{}da} (M'16, SM'19) is the 5G and AI Research Assistant Professor with the Commonwealth Cyber Initiative and a Research Assistant Professor with the Bradley Department of Electrical and Computer Engineering at Virginia Tech. Previously he was a Research Fellow with CONNECT at Trinity College Dublin and a Challenge Research Fellow with Science Foundation Ireland. He is a Fulbright fellow class 2017-18. Jacek's research focuses on modeling and technology design for Next G mobile networks.
\end{IEEEbiographynophoto}

\begin{IEEEbiographynophoto}{Nicola Marchetti} (Senior Member, IEEE) received the Ph.D. degree in wireless communications from Aalborg University, Denmark, and the M.Sc. degrees in electronic engineering and mathematics from the University of Ferrara, Italy, and Aalborg University. He is currently an associate professor in wireless communications with Trinity College Dublin, Ireland. He performs his research under CONNECT, where he leads the Wireless Engineering and Complexity Science (WhyCOM) Lab. His research interests include Autonomous and Self-Organising Networks, Communications for Biology, PHY Algorithms and MAC Protocols, and Quantum Communications and Networks. He has authored in excess of 160 journals and conference papers, and received four best paper awards. He has served as an associate editor of the IEEE Internet of Things Journal and the EURASIP Journal on Wireless Communications and Networks, and is a Fellow of Trinity College Dublin.
\end{IEEEbiographynophoto}

\begin{IEEEbiographynophoto}{Luiz A. DaSilva} is the Executive Director of the Commonwealth Cyber Initiative (CCI), and the Bradley Professor of Cybersecurity at Virginia Tech. He previously held the chair of Telecommunications at Trinity College Dublin, where he served as the Director of CONNECT, the telecommunications research centre funded by the Science Foundation Ireland. Prof. DaSilva's research focuses on distributed and adaptive resource management in wireless networks. Recent and current research sponsors include the National Science Foundation, the Science Foundation Ireland, the European Commission, and DARPA. Prof. DaSilva is a Fellow of the IEEE, for contributions to cognitive networking and to resource management in wireless networks, and has been an IEEE Communications Society Distinguished Lecturer (2015-18) and a Fellow of Trinity College Dublin.
\end{IEEEbiographynophoto}

\end{document}